\def\h{{1\over 2}}

\documentclass[12pt]{article}
\usepackage{latexsym}
\usepackage{enumerate}
\begin{document}
{\baselineskip 22pt

{\ }\qquad\qquad \hskip 4.3in DAMTP/97-96

%{\ }\qquad\qquad \hskip 4.3in hep-th/97xxxx

\vspace{.2in}

\begin{center} {\LARGE Two-Component Reduction of Nahm's 
                       Equations and Hyper-elliptic Solutions}
\\ \baselineskip 13pt{\ }
{\ }\\
Houari Merabet\footnote{Research
supported by the EPSRC grant GR/K50641}\\
 Department of Applied Mathematics \& Theoretical Physics\\
University of Cambridge, Cambridge CB3 9EW, England\\
 M.Houari@damtp.cam.ac.uk\\
\end{center}
\begin{center}
October 1997
\end{center}}

\vspace{10pt}
\begin{quote}
\baselineskip 22pt
\noindent{\bf Abstract} 
We find the general hyper-elliptic solutions to the two-component 
reduced Nahm equations proposed in \cite{HitMan:sym}. 
Elliptic solutions are a special case and can appear only for
specific values of the monopole charges. 
\end{quote}
\baselineskip 22pt
%\baselineskip 28pt
%\vspace{8cm}
%PACS number : 11.10.Lm, 11.27.+d   \\
%Keywords : monopole, gauge theory, Nahm equations, 
%Bogomolnyi equations, elliptic.
\newpage

\section{Introduction}
Monopoles are soliton solutions to the $SU(2)$ Yang-Mills-Higgs
equations in 3 + 1 dimensions \cite{thooft}. In the limit when the Higgs 
potential vanishes, static monopoles are solutions to the Bogomolny 
equations \cite{bogo}
\begin{equation}
F_{ij} = \epsilon_{ijk} D_k \phi 
\label{bogo.eqn}
\end{equation}
where $F_{ij}  =  \partial_i A_j - \partial_j A_i - [A_i, A_j]$
and $D_k \phi  =   \partial_k \phi - [A_k,\phi]$.
The Higgs field $\phi$ and the gauge fields $A_i$ are in the
adjoint representation of SU(2). It has been proven long time ago 
by Nahm \cite{Nahm:con}, that an $N$ charged monopole 
solutions of the Bogomolny equations are uniquely related
to the corresponding solutions of Nahm's equations
\begin{equation}
{d T_k(s)\over ds} ={1\over 2}\epsilon_{ijk} [T_i(s),T_j(s)],
\label{nahm.eqn}
\end{equation}
where the $N\times N$ matrices $T_i$ are meromorphic functions
defined on the interval $[0,2]$, regular on $(0,2)$ with simple
poles at $s=0,2$. The residues at each  pole form an irreducible 
$N$-dimensional representation of $su(2)$. Furthermore,
the  $T_i$ satisfy a reality condition  $T_i(s) =
-T_i\sp\dagger(s) = T_i\sp\dagger(2-s)$. To each solution
of Nahm's equation one can associate a spectral curve defined by
\begin{equation}
P(\eta,\zeta) \equiv \det (\eta + i(T_1 +iT_2)-2iT_3\zeta
-i(T_1-iT_2)\zeta^2) =0
\label{spectral}
\end{equation}
The coefficients of the polynomial (\ref{spectral}) are independent
of $s$ and arise as constant of integrations of Nahm's equations.

Eqs.(\ref{nahm.eqn}) are non-linear first order differential
equations. Although they are more simpler mathematically
than eqs.(\ref{bogo.eqn}) but still only few solutions are known. 
In the present paper, we will show how to obtain general 
solutions for the two-component reduced version of eqs.(\ref{nahm.eqn})
and discover that all the known solutions appear as special cases.

\section{Solutions of 2-component reduced Nahm's equations} 
Let us recall how one obtains two-component reduced Nahm's 
equations \cite{HitMan:sym}. First, one expresses the Nahm
matrices as 
\begin{equation}
T_i(s)  = x(s) \rho_i + y(s) S_i \ .
\label{nahm.mat}
\end{equation}
The real functions $x(s)$ and $y(s)$ have to be regular on $(0,2)$ 
with simple poles at $s=0,2$. The $\rho_i$ form an  $N$-dimensional
representation of so(3),
i.e., $[\rho_i,\rho_j]= 2 \epsilon_{ijk}\rho_k$.
Whereas the $S_i$ are the components of a G-invariant vector 
in $S^{2 k} V$. Here $G$ is a finite subgroup of $SO(3)$. 
The $T_i$ belong to a two-dimensional
space generated by the vetors $\rho_i$ and $S_i$. Therefore 
\begin{eqnarray}
& [S_i,\rho_j]  + [\rho_i,S_j] =  
\epsilon_{ijk} (\alpha \rho_k + \beta S_k) & \\
& [S_i,S_j]   =  \epsilon_{ijk} (\gamma \rho_k + \delta S_k) \ .&
\label{com.mat}
\end{eqnarray}
Nahm's equations reduce  then to
\begin{equation}
{d x\over ds} = 2 x^2 + \alpha x y + \gamma y^2 ;
\qquad  {d y\over ds} = \beta x y + \delta y^2  \ .
\label{dif,sys1}
\end{equation}
Eqs.(\ref{dif,sys1}) represent a two-dimensional system of 
coupled non-linear differential equations. 
One can arrive at the same system 
by considering Nahm's matrices as having values on
the root space of simple Lie algebras \cite{BrzeMer1,BrzeMer2}
\begin{equation}
T_i(s) = \sum_{r\in R} e_{r,i}(s) E_r 
\label{ansatz}
\end{equation}
where $R$ is a set of roots and 
${\bf e}_r = (e_{r,1},e_{r,2},e_{r,3})$ are
three-dimensional vector fields. The matrices $E_{r}$
are the non-commuting generators of a simple Lie algebra.
The reality condition for the Nahm
matrices $T_i$ implies that ${\bf e}_{-r} (s) = - {\bf e}_r^*(s)$ 
while eqs.(\ref{nahm.eqn}) become
\begin{equation}
{d{\bf e}_r(s)\over ds} = \h  \sum_{w\in R}N_{w,r-w} 
{\bf e}_w(s)\wedge {\bf e}_{r-w}(s)
\label{nahm.a.1}
\end{equation}
with constraints
\begin{equation}
\sum_{r\in R} r_\mu {\bf e}_r(s)\wedge {\bf e}_{-r}(s) = 0, 
\qquad \mu =1,\ldots, n.
\label{nahm.a.2}
\end{equation}
where the constants $n$ and $N_{r,w}$ are 
respectively the rank and the structure constants
of a simple Lie algebra. The $(r_1,\ldots,r_n)$ is the root
vector. The two-component reduction (\ref{nahm.mat}) reads 
in terms of the vector fields ${\bf e}_r$ 
\begin{equation}
{\bf e}_r(s) = x(s) {\bf a}_r + y(s) {\bf b}_r \ .
\label{e:2-dim}
\end{equation}
The constant vectors ${\bf a}_r$ and ${\bf b}_r$ can be either
real or purely imaginary and they satisfy
\begin{eqnarray}
&\displaystyle 
 \h  \sum_{s\in R} N_{s,r-s} {\bf a}_s\wedge {\bf a}_{r-s}= 
2 {\bf a}_{r} \ , \qquad
\h  \sum_{s\in R} N_{s,r-s} {\bf b}_s\wedge {\bf b}_{r-s}= 
\gamma {\bf a}_{r} + \delta {\bf b}_{r} & \nonumber \\
& \displaystyle \sum_{s\in R} N_{s,r-s} {\bf a}_s\wedge {\bf b}_{r-s}= 
\alpha {\bf a}_{r} + \beta {\bf b}_{r} \ .&
\label{a.sys}
\end{eqnarray}
The constraints (\ref{nahm.a.2}) read in terms of
${\bf a}_r$ and ${\bf b}_r$
\begin{eqnarray}
\sum_{r\in R} r_\mu ({\bf a}_r \wedge {\bf b}_{-r} 
- {\bf b}_r \wedge {\bf a}_{-r})= 0, \qquad \mu =1,\ldots, n.
\label{a.cons}
\end{eqnarray}
The key idea for solving (\ref{dif,sys1})
is to find a constant linear transformation $u = a x + b y$ and
$v = c x + d y$ such that
\begin{equation}
{d u\over ds} = q_1 u^2 + q_2 v^2 ,
\qquad  {d v\over ds} = q_3 u v .
\label{dif,sys2}
\end{equation}
where $q_1$, $q_2$ and $q_3$ are real constants. 
It is a simple exercise to prove that eqs.(\ref{dif,sys1}) 
can always be put in the form (\ref{dif,sys2}) provided that 
the constants $\alpha, \beta, \gamma$ and $\delta$ satisfy 
some conditions. In the case when $\beta \neq 2$ or $-4$
and $\gamma \neq 0$, one can choose the following transformation
\begin{eqnarray}
& a = (\beta - 2) (\alpha (\beta+2) - 4 \delta) \ ; \qquad   
b = -(\beta-2)^2 (\alpha + 2 \delta) & \nonumber \\
& c = d = (\beta - 2)^2 (\beta + 4) &
\label{trans}
\end{eqnarray}
with the constraint
\begin{equation}
\gamma = \frac{(\beta \delta - 2 \alpha) (\alpha (\beta + 2)
        - 4 \delta)} {(\beta - 2) (\beta + 4)^2} \ .
\end{equation}
Let $p = 2 q_1/q_3$ and $k = 2 q_2/q_3$ then
\begin{eqnarray}
& p = (2 + \beta)/2 ;  \qquad k = (2 - \beta)/2 \nonumber & \\
& q_3 = 4 (\beta - 2) (\alpha \beta + \delta (\beta - 4)) &
\label{p,k}
\end{eqnarray}
Let $t = q_3 s/2$, then the system (\ref{dif,sys2}) reduces to
\begin{equation}
{d u\over dt} = (2 - k) u^2 + k v^2 ,
\qquad  {d v\over dt} = 2 u v \ .
\label{dif,sys3}
\end{equation}
A remarkable feature of this system is that it involves only 
one constant instead of four in eqs.(\ref{dif,sys1}).
The next step is to find a constant of motion for eqs. 
(\ref{dif,sys3}) which will helps us to get rid of one of the 
variables.
\begin{quote}
\textsc{Proposition}.  If $k$ is a positive integer then the 
polynomial $a_k = b_k v^{k-2} (v^2 - u^2)$ of degree $k$ is a constant 
of integration  of eqs.(\ref{dif,sys3}), with $b_k$ an arbitrary constant. 
\end{quote}
It is easy to check that $a_k$ is a constant of integration  
of eqs.(\ref{dif,sys3}) by taking its derivative with respect
to $t$ and then using (\ref{dif,sys3}).
From (\ref{p,k}) we see that the condition
for $k$ to be a positive integer is equivalent to the statement 
that $\beta$ is a negative even integer. 
Since $a_k$ is a constant of integration of Nahm's equations
it should also appear as a coefficient in the spectral curve 
(\ref{spectral}). Therefore the degree $k$ of the polynomial
$a_k$ satisfies $k\leq N$.
The integer $k$ has a simple explanation in the context of 
the root system approach to Nahm's
equations \cite{BrzeMer1,BrzeMer2}. In the case of the $A_n$ root
systems $k = n+1$.\\
Using the expression of $a_k$ in the above proposition, 
we express $u$ as a function of $v$
\begin{equation}
u^2 = v^2 (1 \pm {c^k \over v^k})
\label{u^2}
\end{equation}
where $c^k = |a_k|/|b_k|$. Using (\ref{dif,sys3}), we get a
first-order non-linear differential equation for $v$ 
\begin{equation}
v'^2 = 4 v^4 (1 \pm {c^k \over v^k})
\label{v'^2}
\end{equation}
where $v'=dv/dt$. To solve (\ref{v'^2}) we introduce a function 
$w$ given by $v = c w^{\sigma}$, with $\sigma$ an arbitrary constant.
With this new variable eq.(\ref{v'^2}) reads
\begin{equation}
w'^2 = 4 {{c^2} \over {\sigma^2}} 
(w^{2 (\sigma+1)} \pm w^{(2 - k) \sigma + 2})
\label{w'^2}
\end{equation}
Solutions to eq.(\ref{w'^2}) are of hyper-elliptic and elliptic
types. They are summarized in the following theorem
\begin{quote}
\textsc{Theorem.} 
Hyper-elliptic solutions to eq.(\ref{w'^2})
exist if and only if
\begin{quote}
\begin{enumerate}[(i)]
\item $\sigma = -1$ and $k\geq 5$ or 
\item $\sigma = -{1 \over 2}$ and even $k\geq 8$
\end{enumerate}
\end{quote}
Elliptic solution of Weierstrass type exist for the following cases
only
\begin{quote}
\begin{enumerate}[(i)]
\item $k = 3$ and $\sigma = \pm 1$ 
\item $k = 4$ and $\sigma = \pm {1 \over 2}$ or $\pm 1$ 
\item $k = 6$ and $\sigma = \pm {1 \over 2}$ \ .
\end{enumerate}
\end{quote}
\end{quote}
{\em Proof.} For eq.(\ref{w'^2}) to be an hyper-elliptic
differential equation, the right hand side of (\ref{w'^2})
has to be a polynomial in the variable $w$. In other words,
the constants $l = 2 (\sigma+1)$ and $m = (2 - k) \sigma + 2$ 
have to be non-negative integers.
\begin{quote}
\begin{enumerate}[(a)]

\item If $l = 0$ then $\sigma = -1$ and $m = k$.\\
   The solutions are hyper-elliptic for $k\geq 5$ and elliptic
   for $k=3,4$.

\item If $l = 1$ then $\sigma = -1/2$ and $m = 1 + k/2$.\\
   The solutions are hyper-elliptic for any even $k\geq 8$
   and elliptic for $k=4,6$.

\item If $l = 2$ then $\sigma = 0$ and $m = 2$.\\
   The solutions are logarithmics or constants for any $k$.

\item If $l = 3$ then $\sigma = 1/2$ and $m = 3 - k/2$.\\
   The solutions are elliptic for $k = 4,6$.

\item If $l = 4$ then $\sigma = 1$ and $m = 4 - k$.\\
   The solutions are elliptic for $k = 3,4$.

\item If $l \geq  5$ and $k \geq 3$ the integer $m$ is negative
      and there\\ are no hyper-elliptic solutions. $\Box$
\end{enumerate} 
\end{quote}
The solutions with $k = 3$ and $\sigma = \pm 1$  describe the 
tetrahedral symmetric 3-monopoles as found in \cite{HitMan:sym}, 
whereas $k = 4$ and  $\sigma = \pm {1 \over 2}$ or $\pm 1$ describe 
the octahedral symmetric (4 and 5)-monopoles 
\cite{HitMan:sym,HouSut:oct}. Finally, the cases 
$k = 6$ and $\sigma = \pm {1 \over 2}$ are associated with 
dodecahedral symmetric 7-monopoles \cite{HouSut:oct}.

\section{Conclusions}
We have explicitly solved the two-component reduced Nahm 
equations proposed in \cite{HitMan:sym}. The general solutions
are of hyper-elliptic type. Elliptic solutions of Weierstrass type 
appear only for $k = 3, 4, 6$. The latter correspond to the platonic
symmetric monopoles found by \cite{HitMan:sym,HouSut:oct}.
Hyper-elliptic solutions should describe monopoles with charges
$N \geq 7$, but one has to be sure that they satisfy the right
boundary conditions, i.e., regular on $(0,2)$ with simple
poles at $s=0,2$. The problem then resumes in finding the $N\times N$ 
matrices $\rho_i$ and $S_i$ which statisfy the algebra
(\ref{com.mat}).
This problem can be simplified if one adopts the root system approach
\cite{BrzeMer1}. The problem then consists in solving
the systems (\ref{a.sys}) and (\ref{a.cons}).

{ACKNOWLEDGEMENTS.} I am grateful to Tomasz Brzezi\'nski and
Nick Manton for their encouragements and valuable suggestions.
I wish also to thank Conor Houghton and Sazzad Nasir for their 
critical reading of this manuscript.

\end{document}